\documentclass[runningheads]{llncs}

\usepackage{subcaption}

\usepackage{algorithm}
\usepackage{algpseudocode}
\usepackage{amsmath,amssymb}
\usepackage{graphicx}
\usepackage{booktabs}
\usepackage{multirow}
\usepackage[normalem]{ulem}
\usepackage[font=small,skip=0pt]{caption}
\usepackage{threeparttable}
\usepackage{color}
\usepackage{hyperref}
\usepackage{wrapfig}

\newcommand{\bG}{{\mathbb{G}}}
\newcommand{\bP}[1]{{{\bf P}}\left[{#1}\right]}

\newcommand{\bE}[1]{{\mathbb{E}}\left[{#1}\right]}
\newcommand{\1}[1]{{\bf 1}\left[#1\right]}
\newcommand{\bEg}{{\mathbb{E}}}

\newcommand{\SD}[1]{{\color{blue}#1}}
\newcommand{\TC}[1]{{\color{red}#1}}

\usepackage[textsize=scriptsize]{todonotes}

\newcommand{\fsquare}{\vrule height6pt width7pt depth1pt}   
\newcommand{\myproof}{{\hfill \\ \bf Proof. \ }}           
\newcommand{\myendpf}{\hfill\fsquare \\[0.1in]} 

\newcommand\blfootnote[1]{%
  \begingroup
  \renewcommand\thefootnote{}\footnote{#1}%
  \addtocounter{footnote}{-1}%
  \endgroup
}

\begin{document}

\setlength{\abovedisplayskip}{0pt}
\setlength{\belowdisplayskip}{0pt}

\title{Modeling Citation Trajectories of \\Scientific Papers}

\pdfstringdefDisableCommands{%
  \def\\{}%
}

\author{Dattatreya Mohapatra\inst{1,a} \and
Siddharth Pal\inst{2,b} \and Soham De\inst{3,c}, Ponnurangam Kumaraguru\inst{1,d} \and Tanmoy Chakraborty\inst{1,e}}

\authorrunning{D. Mohapatra et al.}

\institute{IIIT-Delhi, India\\
\email{\{$^a$dattatreya15021,$^d$pk,$^e$tanmoy\}@iiitd.ac.in}\\
\and
Raytheon BBN Technologies, USA\\
\email{$^b$siddharth.pal@raytheon.com}
\and University of Maryland, College Park, USA\\
\email{sohamde@cs.umd.edu}}

\maketitle

\begin{abstract}
Several network growth models have been proposed in the literature that attempt to incorporate properties of citation networks. Generally, these models aim at retaining the degree distribution observed in real-world networks. In this work, we explore whether existing network growth models can realize the {\em diversity in citation growth exhibited by individual papers} -- a new node-centric property observed recently in citation networks across multiple domains of research. We theoretically and empirically show that the network growth models which are solely based on degree and/or intrinsic fitness cannot realize certain temporal growth behaviors that are observed in real-world citation networks. To this end, we propose two new growth models that localize the influence of papers through an appropriate attachment mechanism. Experimental results on the real-world citation networks of Computer Science and Physics domains show that our proposed models can better explain the temporal behavior of citation networks than existing models. 

\keywords{Citation network \and Growth model \and Fitness \and Preferential attachment \and Location-based model.}
\end{abstract}

\section{Introduction}
\blfootnote{The project was partially supported by SERB (Ramanujan fellowship and ECR/2017/00l691) and the Infosys Centre of AI, IIIT Delhi, India.}
\blfootnote{\small The authors would like to thank Dr. Ralucca Gera at Naval Postgraduate School, USA, for initial discussions and insights.}
\blfootnote{This document does not contain technology or technical data controlled under either the U.S. International Traffic in Arms Regulations or the U.S. Export Administration Regulations.}

Over the past two decades, study of citation networks has drawn tremendous attention for various reasons~\cite{abbas2014literature}, such as for 
finding useful academic papers, understanding success of authors, papers and institutes, and  decision making processes like promotion and fund disbursement.

The study of complex networks has emerged as a field to explain nontrivial topological features that occur in a wide range of large networked systems. Citation network is one such example of a complex network, which captures citation relationships between paper sources or documents. A citation network is a directed and acyclic information network, with the documents being the nodes, and directed edges representing citations of one document by another, thereby capturing the flow of information or knowledge in a particular field. An important property of citation networks is the ``in-degree distribution" of nodes.
Several models have been proposed to illustrate this distribution \cite{price1965networks}
-- but while these models aim at retaining the power-law degree distribution in synthetic networks, they fail to reproduce other important properties that real-world citation networks might possess.
For instance, Ren et al. \cite{ren2012modeling} pointed out that existing models underestimate the number of triangles and thus fail to model the high clustering in citation networks, which is closely related to network transitivity and the formation of communities.
In a series of papers \cite{chakraborty2015categorization,chakraborty2018universal}, Chakraborty et al. showed that the temporal growth of the in-degree of nodes ({\em aka}, {\em citation trajectory}) in citation networks can be grouped into five major patterns, 
and such patterns are prevalent across citation networks of different domains, e.g., Computer Science, Physics.

Building on our previous work~\cite{node_visibility}, we first demonstrate that none of the established growth models can adequately realize citation trajectories as observed in the data. This immediately calls for the need to investigate more involved growth models. 

We address this issue by accounting for local and global influences exerted by individual nodes in a network.
We introduce the concept of a `location space' associated with the nodes in the network~\cite{boguna2004models},
and propose two new preferential-attachment growth models based on this concept - Location-Based Model (LBM) and Location-Based Model with Gaussian active subspaces (LBM-G). LBM models both the local and global influences, with new nodes connecting preferentially based on the combined influence. LBM-G is an extension of LBM, which models regions of high activity that can be shifted periodically. We evaluate the proposed  models on both the Computer Science and Physics citation networks. Experimental results show that LBM and LBM-G are indeed more accurate at realizing the citation trajectories of nodes compared to other network models.\footnote{\textbf{Reproducibility:} The codes and the datasets are available at \url{https://github.com/dattatreya303/modeling-citation-trajectories}}

\section{Related Work} Citation networks are growing networks that exhibit certain nontrivial statistical properties. In particular, they have been shown to possess a heavy-tailed power-law in-degree distribution \cite{price1965networks,jeong2003measuring}.
The emergence of a heavy tailed degree distribution was first considered and explained by Price \cite{price1965networks}, who proposed a simple network growth model based on the ``rich-get-richer" mechanism~\cite{price1976general}.
This was later incorporated by Barabasi and Albert~\cite{barabasi1999emergence} and others \cite{krapivsky2005network} in their growth models, that shed light on the concept of preferential attachment.
Following this, a copy mechanism was proposed to describe how a new node could exhibit a mixed behavior by attaching to existing nodes either randomly or based on preferential attachment .
Jeong et al. \cite{jeong2003measuring} found that the attachment probability of a node solely depends on its degree, once again confirming preferential attachment.
The evolution of citation networks was also modeled by preferential attachment with a time dependent initial attractiveness \cite{bianconi2001competition,eom2011characterizing,nguyen2012fitness}.
Their model reproduces certain growth of in-degree of a paper right after its publication, called `citation bursts'.
Peterson et al. \cite{peterson2010nonuniversal} further characterized this phenomenon. 
Other models combine the influence of the degree of a node with its intrinsic fitness in defining the network growth dynamics~\cite{bianconi2001competition,nguyen2012fitness}.

Another class of growth models employ decay factors to model the temporal nature of certain node properties\cite{aging_dorogovtsev}.
Dorogovtsev et al. \cite{aging_dorogovtsev} conducted the initial study on such models and found that the networks generated did retain the scaling behavior for a certain range of aging exponent.
Medo et al. \cite{temporal_effects} further combined decay with heterogeneous fitness values to show that employing decay is vital to retain scaling behavior in preferential attachment models.
There was also a need to explore the effect of the aging factor on other global properties of the network, as exhibited by Zhu et al. \cite{effect_aging} and Medo et al. \cite{temporal_effects} studied the average node distance and clustering behavior under the effect of aging factor, and found that network underwent a continuous transformative process which could be controlled by tuning the aging exponent.

A growing body of research focused on the temporal growth of the in-degree of nodes (citation trajectory) in citation networks. Motivated by Xie et al. \cite{xie2015modeling}, 
Chakraborty et al. \cite{chakraborty2015categorization} studied a large citation network and categorized citation trajectory of nodes into five distinct patterns depending on the number and position of peaks in the trajectory. They further showed that the citation trajectory of authors also follows these five patterns and unfolded many interesting causes responsible for these patterns  \cite{chakraborty2018universal}. We posit that a new network growth model is needed to explain these patterns, and our propositions in this work are a means to that end.

\section{Dataset}
\label{sec:datasets}
Publication datasets of two domains (Computer Science and Physics) are used in this paper.

\begin{table}
\centering
\caption{Statistics of the datasets.}
\label{tab:dataset}
\scalebox{0.9}{
\begin{tabular}{|c|c||c|c|}
\hline
\multicolumn{2}{|c||}{{\bf MAS}} & \multicolumn{2}{c|}{{\bf APS}}\\ 
\hline
{\bf \# of nodes} & {\bf \# of edges} & {\bf \# of nodes} & {\bf \# of edges} \\\hline
\multicolumn{4}{|c|}{\bf Complete network (1960-2010)}\\\hline
711,810 & 1,231,266 & 463,347 & 4,710,547\\\hline
\multicolumn{4}{|c|}{\bf Network under examination (1960-2000)}\\\hline
282,919 & 589,201 & 277,999 & 2,474,076 \\\hline
\multicolumn{4}{|c|}{\bf Seed network}\\\hline
4,134 & 4,872 & 3,569 & 4,108 \\\hline
\end{tabular}}
\end{table}

\textbf{Microsoft Academic Search (MAS):} We used the publication dataset of Computer Science domain released by Microsoft Academic Search (MAS)~\cite{ref:mas}.
We further filtered a subset of this dataset for experimental purposes, considering only papers published till the year 2000 because all the papers under examination require at least 10 years of citation history as suggested in  \cite{chakraborty2015categorization,chakraborty2018universal} to obtain the citation growth trajectory. 

\textbf{APS Journal Data:} 
The American Physical Society (APS) journal dataset\footnote{https://journals.aps.org/datasets} contains articles published in  journals like Physical Review Letters, Physical Review and Reviews of Modern Physics publications. Similar to MAS, we only considered articles published between 1960 and 2000.

All the models used in this papers start with a seed network, consisting of papers published in 1960-1975.
Table \ref{tab:dataset} summarizes these datasets. 

\section{Preliminaries}

\subsection{Categorization of Citation Trajectories}
\label{sec:category}
\begin{figure}[t]
\centering
\includegraphics[scale=0.5]{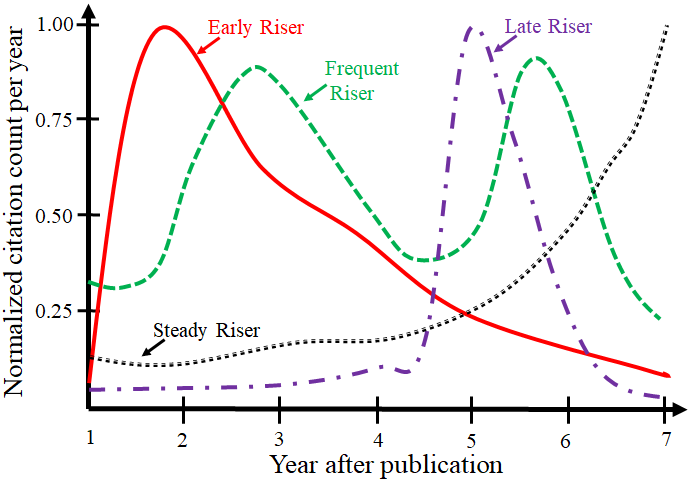}
\caption{Toy example of citation trajectories. The `Others' category does not follow any consistent pattern and is therefore not shown here.}
\label{fig:categories}
\end{figure}

Chakraborty et al. \cite{chakraborty2015categorization,chakraborty2018universal} defined `citation trajectory of a paper' as the (non-cumulative) number of citations (normalized by maximum citation count at any year) per year the paper receives till the time of analysis. One can consider a citation trajectory as a time-ordered set of data points (integers).
They observed that contrary to the general consensus that the shape of the citation trajectory of papers are the same \cite{barabasi1999emergence,jeong2003measuring}, there are five different shapes of citation trajectories prevalent across different domains of citation network (see Figure \ref{fig:categories} for the toy examples of trajectories):
(i) \textbf{Early Risers (ER)} with a single  peak in citation count within the first five years (known as {\bf activation period}) after publication; (ii) \textbf{Frequent Risers (FR)} with distinct multiple peaks; (iii) \textbf{Late Risers (LR)} with a single peak after five years of publication (but not in the last year); (iv) \textbf{Steady Risers (SR)} with monotonically increasing citation count; and (v) \textbf{Others (OT)} whose average numbers of citations per year are less than 1. 

\subsection{Citation Networks as Random Growth Models}

The class of random growth models is a natural choice for analyzing citation networks owing to their growing nature.
Network growth models define a sequence of random graphs $\{\bG_t, \ t=0,1,\ldots \}$, where $\bG_t = (V_t,\bEg_t)$, with $V_t$ and $\bEg_t$ being the set of nodes and edges in $\bG_t$, respectively. In a network growth model, we have 
$V_{t-1} \subset V_{t}$ and $\bEg_{t-1} \subseteq \bEg_{t}$ 
for every $t=0,1,\ldots$. 
Thus, new nodes arrive at every time step $t$, and form connections with existing nodes randomly, thereby adding to the edge set of the previous graph $\bG_{t-1}$. The degree of node $i$ in graph $\bG_t$ is denoted by $D_t(i)$. Table \ref{tab:terminology} defines some additional notations used in this paper.

\begin{table}[!t]
\centering
\caption{Important notations and denotations.}
\label{tab:terminology}
\begin{tabular}{|c||l|}
\hline
{\bf Symbol} & {\bf Definition} \\
\hline
$D_t(i)$ & Degree of node $i$ at time $t$ \\ 
$\chi_i$ & Location of node $i$ \\ 
$L$ & Location space (i.e., $\chi_i \in L$) \\ 
$\xi_i$ & Fitness of node $i$ \\
$F$ & Fitness space (i.e., $\xi_i \in F$) \\ 
$p_i(t)$ & Attachment prob. to node $i$ at  $t$\\
\hline
\end{tabular}
\vspace{-5mm}
\end{table}

\subsection{Popular Classes of Network Growth Models}
\label{sec:existing}
We want to investigate whether some popular classes of network growth models can exhibit the different citation trajectories mentioned in Section \ref{sec:category}. To this end, we first look at three popular classes of network growth models.

\textbf{Barab\'asi-Albert (BA) Model:}
The BA model \cite{barabasi1999emergence} is based on the principle of ``preferential attachment", where new nodes connect preferentially to existing nodes with higher degree. The probability that a new node at time $t+1$ connects to node $i \in V_t$ is given by,
\begin{equation}
\label{ap:ba}
p_i ^{BA} (t+1) = \frac{D_t(i)}{\sum _{j \in V_t} D_t(j) }.
\end{equation}
Since new nodes have low initial degrees, owing to initial in-degree  being zero, subsequent nodes that join the network will attach to older nodes with higher probability. Hence, it is evident that the BA model cannot simulate the initial growth in citations after publication, i.e., Early Risers, which is often observed in citation networks (See Table \ref{tab:existing}).  This motivates us to incorporate the intrinsic `attractiveness potential', or `fitness', of a paper in the model.
Section \ref{sec:theory} gives theoretical justifications for the same.

\textbf{Fitness-based Models:}
As argued previously, to model initial peak in the citation trajectory of papers, we associate a fitness value~\cite{fitness_intrinsic,fitness_vertex_intrinsic,fitness_generative} with each node. We assume an independently and identically distributed (iid) sequence of random variables (rvs) $\{ \xi, \xi _i, \ i=1,2,\ldots\}$ randomly drawn from a power law distribution $\xi$, e.g., Pareto distribution, with $\xi _i$ denoting the fitness of node $i$. 
In the additive model the attachment probability of a node is directly proportional to the sum of its degree and fitness; while in the multiplicative model it is directly proportional to their product. The attachment probabilities that new nodes connect to node $i \in V_t$ at time $t+1$ are given by
\begin{align*}
\text{Additive: } \qquad & p_i ^{AF} (t+1) = \frac{D_t(i)+\xi_{i}}{\sum _{j \in V_t} (D_t(j)+\xi_{j})}, \\
\text{Multiplicative: } \quad & p_i ^{MF} (t+1) = \frac{D_t(i) \cdot \xi_{i}}{\sum_{j\in V_t} D_t(j) \cdot  \xi_{j}}.
\end{align*}

As expected, the fitness-based models are able to achieve some amount of initial citation growth after the publication of a paper (See Table \ref{tab:existing}).
Furthermore, the multiplicative model can achieve a non-zero fraction of Steady Risers category in the APS citation network, which is realized to a lesser extent in the BA model or the additive model.  Section \ref{sec:theory} presents theoretical justifications.

\begin{table}[t]
\caption{Percentage of papers in different categories present in two real datasets -- (a) MAS and (b) APS, and that generated by the existing models (BA: Barab\'asi-Albert model, Add: Additive-fitness model, Mult: Multiplicative-fitness model) as well as our models (LBM: Location-based model, LBM-G: Location-based model with Gaussian active subspace) with best parameter setting. We also measure the (square of) Jensen-Shannon distance (JSD) \cite{jensenShannonDist} of the results for each model w.r.t. the percentage obtained in real datasets (see Section \ref{sec:eval} for more details).}
\label{tab:existing}
\centering
\begin{tabular}{c}
     \includegraphics[width=\columnwidth]{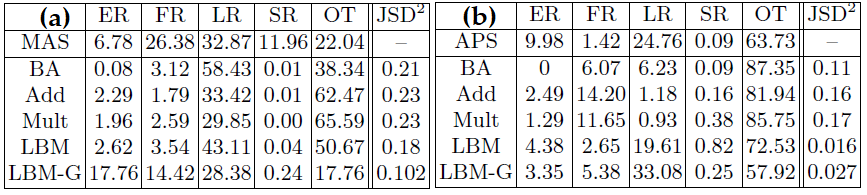}
\end{tabular}
\vspace{-5mm}
\end{table}

\section{Theoretical Analysis}
\label{sec:theory}
The following theorem 
describes the change in attachment probabilities of nodes over time for the three growth models described in Section \ref{sec:existing}.
We define $\Xi _t = \sum _{i \in V_t} \xi _i$
and $\psi _t = \sum _{i \in V_t} \xi _i D_t(i)$, for 
$t=0,1,\ldots$. For clarity in presentation,
we assume that a single node enters at any time step $t$, and forms a connection with one node in the 
existing graph $\bG_{t-1}$. We label the incoming node by the time index of its entry to the network, i.e., $V_t = \{0,1,\ldots,t\}$ for $t = 0,1,\ldots$. All our results can be easily extended to more general scenarios where multiple nodes enter the network, and incoming nodes form multiple connections.

\begin{theorem}
\label{main_thm}
For every $t=0,1,\ldots,$ and $i$ in $V_{t-1}$: Let $\bG _{t-1}$ be the graph at time $t-1$. Then the expected changes in attachment probabilities of node $i$ are given below.
\begin{itemize}
\item[(i)] Barab\'asi-Albert (BA) model:
\begin{equation}
\bE{ p_i ^{BA} (t+1) - p_i ^{BA} (t) | \bG _{t-1} } = -\frac{D_{t-1}(i)}{4t(t-1)}
\label{eq:Change-BA-expected}
\end{equation}
\item[(ii)] Additive fitness (AF) model:
\begin{equation}
\bE{ p_i ^{AF} (t+1) - p_i ^{AF} (t) | \bG _{t-1}, \xi_t } = -\frac{\left( \xi _i +D_{t-1}(i) \right) \left( \xi _t + 1 \right)}{ \left( \Xi _{t-1} + 2(t-1) \right) \left( \Xi_t + 2t \right) }
\label{eq:Change-AF-expected}
\end{equation}
\item[(iii)] Multiplicative fitness (MF) model:
\begin{equation}
\bE{p_i^{MF}(t+1) - p_i^{MF}(t) | \bG_{t-1}} \gtrsim \xi_i D_{t-1} (i) \frac{\sum _{j \in V_t} \xi_j D_t(j) \left[ \xi _i - \xi_t - \xi_j \right]}{\psi _{t-1}^2 (\psi _{t-1} +\xi _i + \xi_t)}
\label{eq:Change-MF-expected}
\end{equation}
\end{itemize}
\label{lemma:Change_in_visibility}
\end{theorem}

\myproof Fix $t=0,1,\ldots$, and $i$ in $V_t$.\\
\textbf{Preferential Attachment Model:}
The difference in the attachment probability of node $i$ in the BA model between time $t+1$ and $t$ is given as
\begin{align}
    p_i ^{BA}(t+1) - p_i^{BA}(t) = 
    \frac{D_t(i)}{2t} - \frac{D_{t-1}(i)}{2(t-1)}
    = \frac{D_{t-1}(i)+\1{S_t = i}}{2t} - \frac{D_{t-1}(i)}{2(t-1)}
    \label{eq:proof-BA-Change-1}
\end{align}
 Furthermore, by noting that when looking at the expected difference in attachment probability conditioned on the graph at time $t-1$, $S_t$ is the only random variable in 
\eqref{eq:proof-BA-Change-1}, we obtain
\begin{align}\small
    \bE{ p_i ^{BA} (t+1) - p_i ^{BA} (t) \ | \ \bG _{t-1} } 
    &= \frac{D_{t-1}(i)+\text{Pr}[S_t = i \ | \ \bG_{t-1}]}{2t} - \frac{D_{t-1}(i)}{2(t-1)}
    \nonumber \\
    &= \frac{D_{t-1}(i) + \frac{D_{t-1}(i)}{2(t-1)}}{2t} - \frac{D_{t-1}(i)}{2(t-1)} \nonumber 
\end{align}
and \eqref{eq:Change-BA-expected} follows.\\
\textbf{Additive Fitness Model:}
Similarly in the additive fitness model, the difference in the attachment probability of node $i$ can be 
written as
\begin{align}\small
    p_i^{AF}(t+1) - p_i ^{AF}(t) 
    &= \frac{\xi _i + D_t(i)}{ \sum _{j \in V_t} \left( \xi _j + D_t(j) \right)  }  - 
    \frac{\xi _i + D_{t-1}(i)}{ \sum _{j \in V_{t-1}} \left( \xi _j + D_{t-1}(j) \right) }
    \nonumber \\
    &= \frac{\xi _i + D_{t-1}(i)+ \1{S_t = i} }{ \Xi_{t-1} + \xi _t + 2t  }  - 
    \frac{\xi _i + D_{t-1}(i)}{ \Xi_{t-1} + 2(t-1)  }.
\end{align}
Taking expectation on both sides
conditioned on $\bG_{t-1}$ and $\xi_t$ leads to \eqref{eq:Change-AF-expected}.\\
\textbf{Multiplicative Fitness model:}
The difference in the attachment probability of node $i$ can be written for the multiplicative model as follows
\begin{align}\small
    p_i ^{MF}(t+1) - p_i ^{MF} (t) 
    &= \frac{\xi _i D_t(i)}{\sum _{j \in V_t} \xi _j D_t(j)} - 
    \frac{\xi _i D_{t-1}(i)}{\sum _{j \in V_{t-1}} \xi _j D_{t-1}(j)}
    \nonumber \\
    &= \frac{\xi _i \left[ D_{t-1}(i) + \1{ S_t = i} \right]}{\psi _{t-1} + \xi_{S_t} + \xi _t} 
    - \frac{\xi _i D_{t-1}(i)}{\psi _{t-1}},
    \label{eq:MF-model-intermediate-1}
\end{align}
where \eqref{eq:MF-model-intermediate-1} follows by noting that $\psi _t = \psi _{t-1} + \xi _{S_t}+\xi_t$, because only node $S_t$ gets a new edge from $\xi_t$ and its fitness degree product therefore only increases by its own fitness $\xi_{S_t}$.\\
Furthermore, we lower bound the expected change 
in visibility as follows
\begin{align}
    &\bE{ p_i ^{MF}(t+1) - p_i ^{MF}(t) | \bG_{t-1}, \xi _t} 
    \nonumber \\
    &= \xi_i \left[ \frac{\psi_{t-1} \bP{S_t = i | \bG_{t-1}, \xi_t } }
    { \psi_{t-1} \left( \psi_{t-1} + \xi_i + \xi_t \right) } 
    - \frac{D_{t-1}(i)}{\psi_{t-1}(j)} \left[ \sum_{\ell \in V_{t-1} }
    \bP{ S_t = \ell | \bG_{t-1}, \xi_t} \cdot \frac{\xi_\ell + \xi _t}{ \psi _{t-1} + \xi_\ell + \xi_t }
    \right] \right]
    \nonumber \\
    &\geq \xi_i \left[ \frac{ \xi_i D_{t-1}(i) }
    { \psi_{t-1} \left( \psi_{t-1} + \xi_i + \xi_t \right) } 
    - \frac{D_{t-1}(i)}{\psi_{t-1}} \cdot 
    \frac{\sum_{ \ell \in V_{t-1}} \xi_\ell D_{t-1}(\ell) (\xi_\ell + \xi_t ) }{\psi_{t-1} ^2}\right]
    \nonumber
\end{align}
\begin{align}
    &= \frac{\xi_i D_{t-1}(i)}{\psi_{t-1}} 
    \left[ \frac{\xi_i \psi_{t-1}^2  - \sum_{\ell \in V_{t-1}} \xi_\ell D_{t-1} (\ell) (\xi_\ell +\xi_t) 
    [\psi_{t-1} + \xi_i + \xi_t ]}{ \psi_{t-1}^2 (\psi_{t-1} + \xi_i + \xi_t) }\right]   
    \nonumber \\
    & \simeq \xi_i D_{t-1}(i)\left[ \frac{\xi_i \psi_{t-1} - \sum_{ \ell \in V_{t-1}} \xi_\ell D_{t-1} (\ell) (\xi_\ell +\xi_t) }{ \psi_{t-1}^2 (\psi_{t-1} + \xi_i + \xi_t) }\right]
    \nonumber
\end{align}
and the result follows.
\myendpf

Theorem \ref{lemma:Change_in_visibility} indicates that for both the BA model and the additive fitness model, the probability of attachment will reduce over time in expectation for all the nodes. However, for the multiplicative model the probability of attachment could increase over time if the node has sufficiently high initial fitness $\xi_i$. 
This gives a strong theoretical justification for using multiplicative model for realizing Steady Risers, in that nodes with high initial fitness could possibly maintain or increase their attachment probability over time. However in Table \ref{tab:existing}, we observe that the proportion of Steady Risers is negligible in comparison to the MAS data. This is probably because only a few nodes get the opportunity to be Steady Risers in the multiplicative model.

\section{Proposed Models}
\label{sec:model:proposed}
We propose two models which introduce the concept of a `location space' to model the local visibility of nodes in a network and capture different citation growth patterns, while keeping the global influence due to degree and fitness in a multiplicative fashion. This allows us to overcome the limitation of low number of Steady Risers in the multiplicative model by restricting the influence of nodes to their locality.

\subsection{Location-based Model (LBM)}
\label{sec:model:LBM}
Popular articles in scientific networks are prominent in their subfield of interest (say, Machine Learning in Computer Science domain), and do not exert the same amount of influence across other subfields.
We overcome this limitation in the fitness models by restricting the region of influence for incoming nodes. 

The location-based growth model (LBM) proposed here captures the notion of `local influence' of a node in the network. Each node is assigned a location vector drawn from a distribution over the location space $L$. This location vector serves as a representation of the subfield to which a research article belongs. Given a sequence of iid location vectors \{$\chi, \chi_t, \ t = 0, 1, ...$\} and fitness vectors \{$\xi, \xi_t, \ t = 0, 1, ...$\}, with the location vectors being uniformly distributed over the location space $L$, and the fitness vectors being Pareto distributed, the attachment rule is given as,
\begin{equation}
\label{eq:ap_location}
p_i ^{LBM}(t+1) = e^{-\gamma d(\chi_i, \chi_{t+1})} \cdot \xi _i \cdot D_t(i)
\end{equation}
where, $d(\cdot, \cdot)$ can be any distance metric, and $\gamma$ is a decay factor governing how fast the attachment probability decays with distance in the location space.
We use Euclidean distance metric $d(\cdot, \cdot)$ and report results for different values of $\gamma$. Algorithm 2 describes the exact formulation of the network growth using LBM.

\begin{figure}[!t]
    \centering
    \includegraphics[width=\columnwidth]{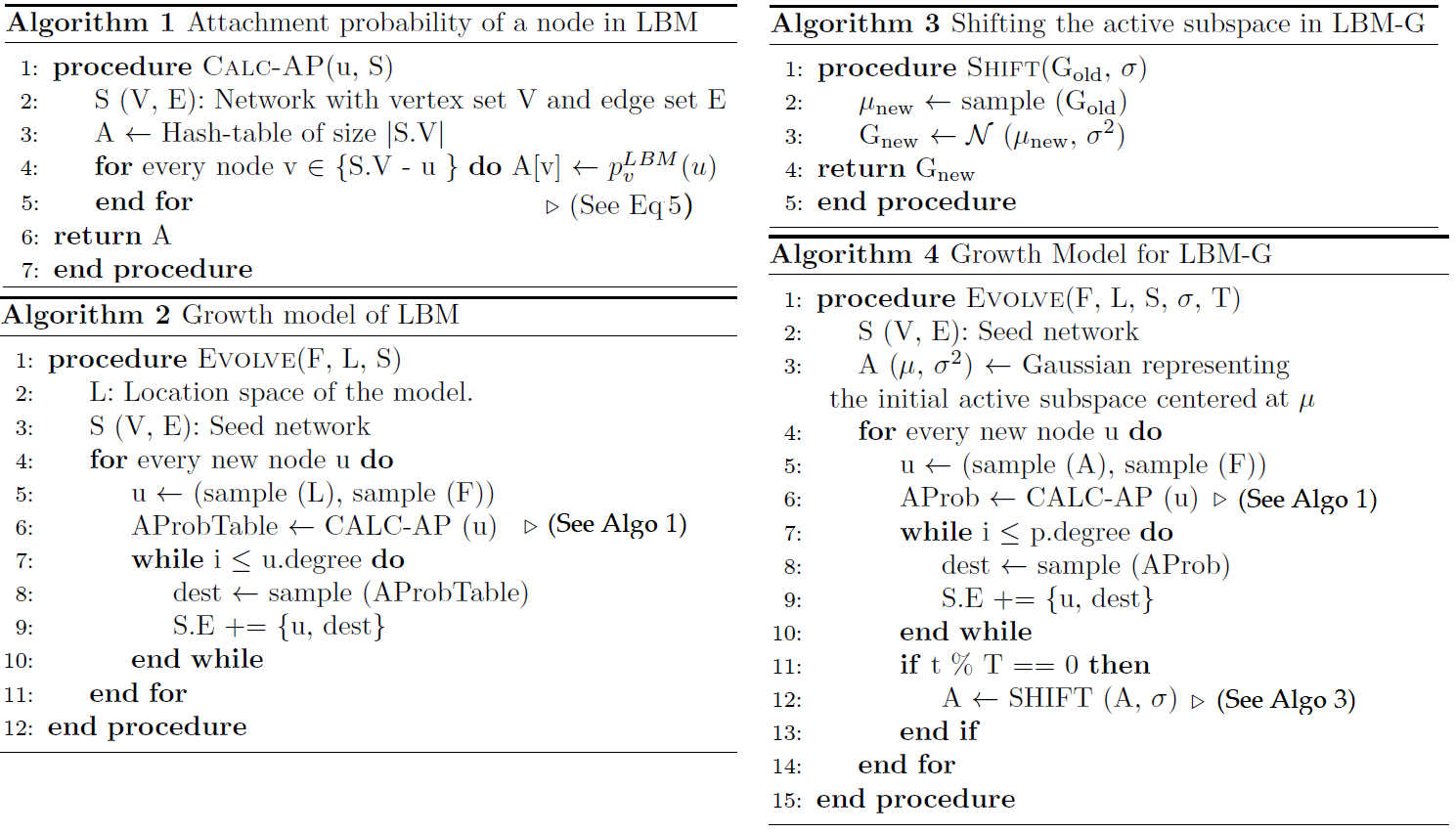}
    \label{fig:lbm_algos}
    \vspace{-9mm}
\end{figure}

\subsection{LBM using Gaussian Active Subspaces (LBM-G)}
\label{sec:model:gaussian}
In LBM, sampling a node location over the entire space $L$ implies a faulty scenario where new papers entering the network at the same time step seemingly belong to widely different subfields -- whereas we generally observe heightened reasearch activity in only a handful of subfields at a time. The growth model should take these regional spikes into account while assigning node locations and also incorporate their shifting nature.

We introduce the concept of \textit{active subspaces} which generates the locations for new papers entering the network. We realize these active subspaces through multi-dimensional Gaussian distributions over the location space $L$. For simplicity, in this paper we only assume one active subspace at every time step. The location vector of node $t$, $\chi _t$, is chosen as $\chi_t \sim \mathcal{N}(\mu _t , \sigma ^2)$, where $\mu _t$ and $\sigma ^2$ are the mean and variance of the Gaussian at time $t$.

The Gaussian distribution is then updated either after entry of fixed number of nodes, or after certain number of times steps. For ease of exposition, we assume that single node enters at every time step. Under this assumption, both the update techniques are identical, which is afterwards relaxed in Section \ref{sec:eval}. After entry of $S$ nodes or $S$ time steps, the mean of the Gaussian distribution is updated as follows
\begin{equation}
\mu _{\ell+1} \sim \mathcal{N}(\mu _\ell , \rho ^2), \ \ell = kS \mbox{ and } k=1,2,\ldots
\end{equation}
where, $\rho^2$ captures the variance in the shift of the active subspace. Once the location vectors are drawn, the attachment rule is identical to the LBM model. The pseudo-code of LBM-G is shown in Algorithm 4.

\begin{table}[t]
\caption{Percentage of papers in different categories obtained from the simulated network of LBM for (a) MAS and (b) APS. We also vary $\gamma$ and report the result as well as the similarity with the real data in terms of JSD.}
\label{tab:lbm}
\centering
\includegraphics[width=\columnwidth]{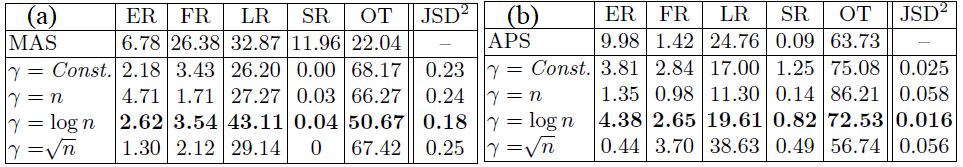}
\vspace{-7mm}
\end{table}

\section{Experimental Results}
\label{sec:eval}
We adopt the experimental setup proposed by Chakraborty et al. \cite{chakraborty2015categorization} to identify the proportion of nodes belonging to each citation trajectory.

{\bf Simulation Setup:} Each model takes a seed network as input 
(see Table \ref{tab:dataset} for the statistics) and grows the network according to its attachment rules. For fair comparison with the real network, we take the information of number of papers published in each year $t$ 
and the out-degree (number of references) of each paper from the real-world dataset. This ensures that the resultant network obtained from each model will have exactly equal number of nodes and edges as of the real network. In particular, the simulation for each model consists of the following steps: 
\begin{enumerate}
\item For every year $t$, we insert equal number of papers (nodes) as present in the real network. We also keep the same out-degree of the papers as that present in the real network.
\item For a new node, let us assume $k$ to be its out-degree. The model will select $k$ nodes randomly without replacement from the existing network. The weight of each node in the network is defined by the attachment rule of the model. The model will then connect the new node to each of these selected $k$ nodes. 
\item We increment the year from $t$ to $t+1$ and repeat Steps 1 - 2.
\end{enumerate}
The remaining part of the section will present the similarity of the simulated network obtained from each model with the real network based on how efficiently the model captures the paper distribution in each category.
We use (the square of) the Jensen-Shannon distance (JSD)  \cite{jensenShannonDist} to measure the similarity between two distributions\footnote{\small Other distribution similarity measures such as RMSE may not be appropriate here due to the varying orders of magnitude of difference between categories. For instance, RMSE would prefer that parameter setting which produces a biased distribution (i.e., one category with high proportion of nodes).}. Jensen-Shannon Distance between two distributions $P$ and $Q$ is defined as, $JSD (P || Q) = \sqrt[]{\frac{1}{2}D (P || M) + \frac{1}{2}D (Q || M)}$, where $M = \frac{P + Q}{2}$. $D (P || M)$ represents the Kullback-Leibler Divergence between the probability distributions $P$ and $M$. Lower the value of JSD, higher the similarity. We report the values of $JSD^2$ with the results as it makes differentiating the effect of two sets of parameters easier.

Note that although the simulation allows all papers published till 2010, for the evaluation we only consider the category of those papers published till 2000 as we need at least 10 years of its citation history to identify the citation trajectory of each paper \cite{chakraborty2015categorization,chakraborty2018universal} (see Table \ref{tab:dataset} for the statistics).

\subsection{Results of LBM}\label{sec:resultLBM}
We analyze the network obtained from the LBM model and measure the proportion of papers in each category of citation trajectory. We also observe  the effect of the scalar factor $\gamma$. The following regimes of $\gamma$ are considered:
(a) constant scalar factor $\gamma$;
(b) linear dependency on the number of nodes $n$, $\gamma = n$;
(c) sub-linear dependency on the number of nodes $n$, $\gamma = n^{0.5}$;
(d) logarithmic dependency on the number of nodes $n$, $\gamma = \log n$. 
Table \ref{tab:lbm} reports the JSD \cite{jensenShannonDist} between the paper category distribution of the LBM model and the real network. We notice that for both MAS and APS datasets, the best result (minimum JSD) is obtained with $\gamma=\log n$. Therefore, we choose $\gamma=\log n$ as default for LBM. 

\begin{table}[!t]
    \centering
    \caption{Percentage of papers in different categories obtained from the simulated network of LBM-G for (a) MAS and (b) APS. We also vary $S$  ({\bf in terms of months}) and $\sigma$ (standard deviation) and report the result as well as the similarity with the real data in terms of JSD.}
\label{tab:gaussian_MAS_APS_months}         \includegraphics[width=\columnwidth]{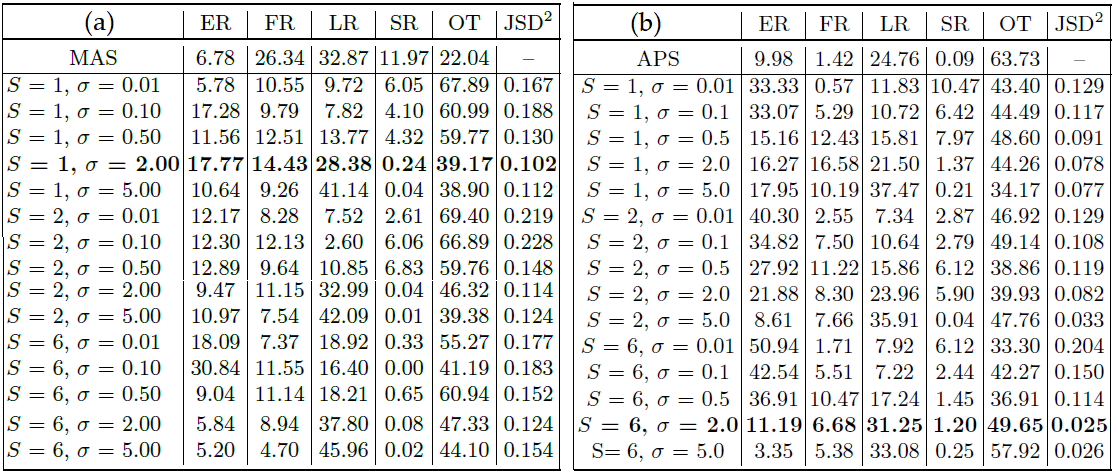}
\vspace{-7mm}
\end{table}

\subsection{Results of LBM-G}\label{sec:resultLBMG}
Along with $S$ indicating the frequency at which LBM-G model shifts the active subspace, we have another hyper-parameter, $\sigma$, the standard deviation of each Gaussian distribution representing a subspace. We consider two different strategies of shifting the active subspace: in terms of months (Table \ref{tab:gaussian_MAS_APS_months}), and the number of nodes (Table \ref{tab:gaussian_MAS_APS_nodes}).

We obtain the lowest JSD for $S = 1$ month and $\sigma = 2.0$, i.e., the active subspace is shifted every month and the new Gaussian subspace is chosen from a Gaussian distribution with standard deviation 2. However, the more interesting pattern to observe in Table \ref{tab:gaussian_MAS_APS_months} is that we consistently get good results (low JSD) with high $\sigma$ values, while keeping the frequency of shifting the active subspace constant. This stands true for the APS dataset as well. We set $S=1$, $\sigma=2$ and the frequency of shifting the active subspace in terms of months as default for LBM-G. 


We can also vary $S$ and $\sigma$ to control the proportion of nodes belonging to each category at a coarser level. We observe the following patterns in Tables \ref{tab:gaussian_MAS_APS_months} and \ref{tab:gaussian_MAS_APS_nodes}:

\begin{itemize}
\item The proportion of Late Risers has a direct correlation with $\sigma$. This is probably because nodes belonging Late Risers acquire a peak after the activation period. A high $\sigma$ makes it more likely that the active subspace shifts further away from the node, thus delaying re-activation of the node's location.
\item The proportion of Early Risers is seen to have a positive correlation with $S$, in general. This can be explained by the effect $S$ has on the activation period of a node. A higher value of $S$ means the node's location stays activated for a longer period of time, thus giving it more opportunities to achieve a peak.
\item The proportion of Steady Risers has a negative correlation with $S$, in general. One requirement for a node belonging to the Steady Risers category is that it should not achieve a peak during the activation period. This means that the factor controlling the proportion of Early Risers can be used to control the proportion of Steady Risers as well. A lower value of $S$ means the active subspace gets shifted frequently, thus not giving ample opportunity to form a very high peak. A higher value of $S$ would instead keep the location activated for an extended period of time, which may lead to higher peaks early on when the number of nodes in the network is low, and then lower peaks as the number of nodes increases.
\item The proportion of Frequent Risers should depend upon the frequency with which a certain subspace is re-activated. We observe that this proportion generally peaks for moderate values of $S$ and $\sigma$.
\end{itemize}

\begin{table}[!t]
    \centering
    \caption{Percentage of papers in different categories obtained from the simulated network of LBM-G for (a) MAS and (b) APS. We also vary $S$ ({\bf in terms of nodes}) and $\sigma$ (standard deviation) and report the result as well as the similarity with the real data in terms of JSD.}
    \label{tab:gaussian_MAS_APS_nodes}
         \includegraphics[width=\columnwidth]{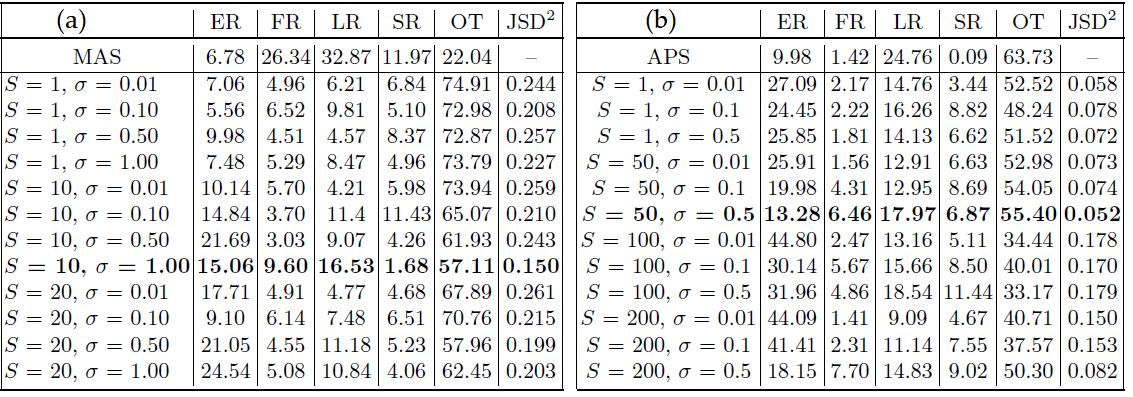}
         \vspace{-8mm}
\end{table}

\begin{figure*}
\includegraphics[width=\linewidth]{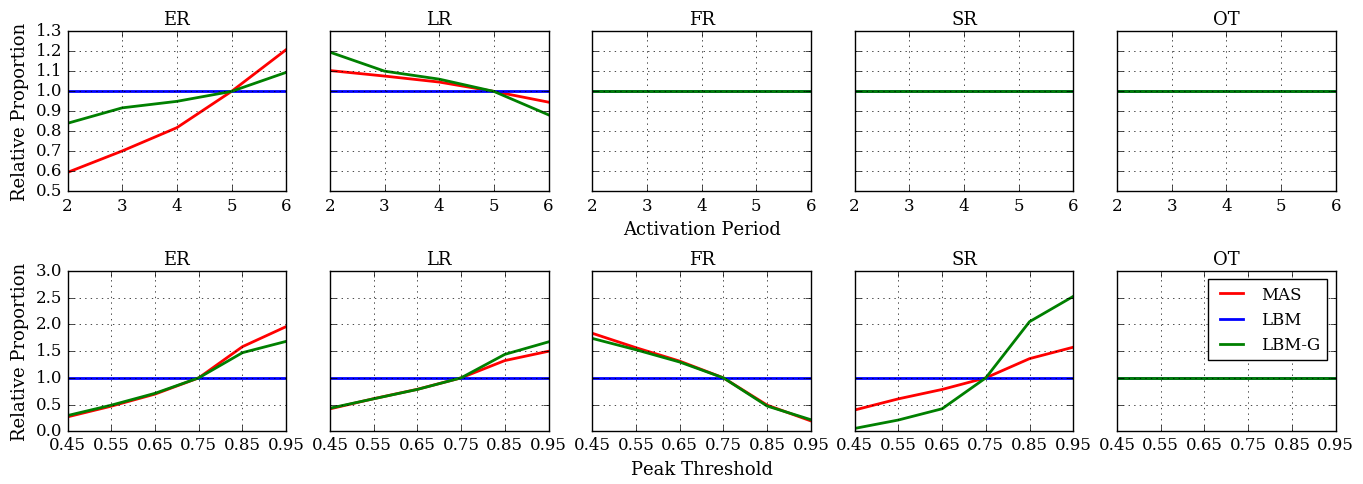}
\caption{(Color online) [MAS dataset] Sensitivity of trajectory classifier on two parameters -- activation period and peak threshold. We vary activation period from 3-7 years (5 years as default) and peak threshold from $0.3-0.9$ ($0.6$ as default) and measure the relative  proportion of different categories w.r.t. that with the default setting on the MAS dataset (red) and the simulated networks obtained from LBM (blue) and LBM-G (green) with default model parameters. We observe that the relative change does not vary much w.r.t. the default setting. In most of the plots, three lines were superimposed (e.g., bottom, right most subplot).}
\label{fig:mas_robust}
\end{figure*}

\begin{figure*}[!t]
\includegraphics[width=\linewidth]{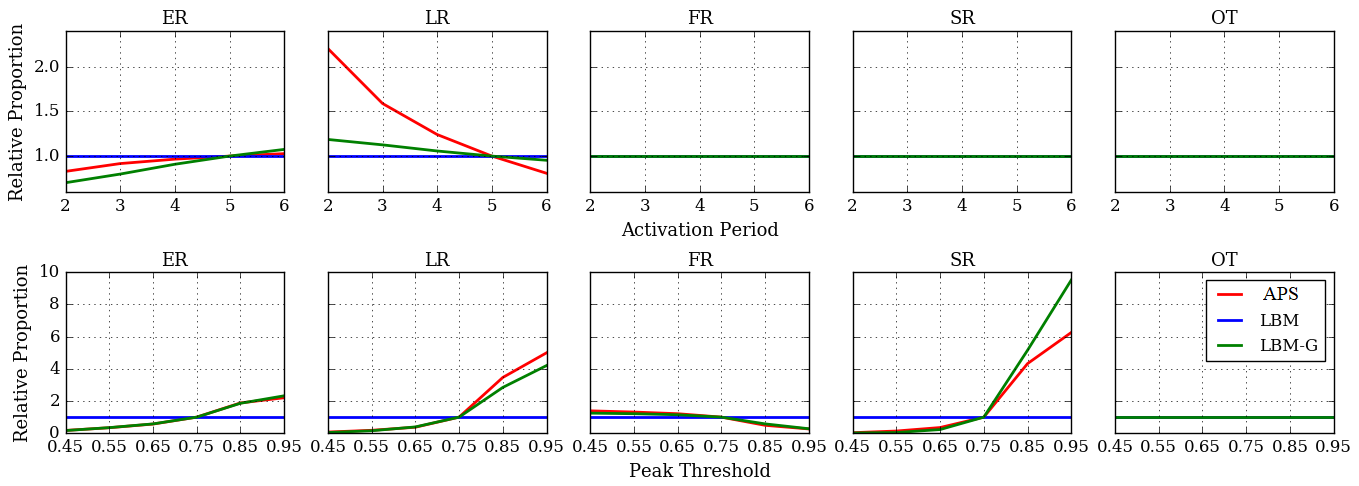}
\caption{(Color online) [APS dataset] Sensitivity of trajectory classifier on two algorithmic parameters -- activation period and peak threshold. We vary activation period from 3-7 years (5 years as default) and peak threshold from $0.3-0.9$ ($0.6$ as default) and measure the relative  proportion of different categories w.r.t. that with the default setting on the MAS dataset (red) and the simulated networks obtained from LBM (blue) and LBM-G (green) with best model parameters. We observe that the relative change does not vary much w.r.t. the default setting. In most of the plots, three lines were superimposed.}
\label{fig:aps_robust}
\end{figure*}

\subsection{Sensitivity Analysis}\label{sec:sensitiity}
The results reported throughout the paper are obtained by setting activation period to 5 years and peak threshold to 0.75 as default for trajectory categorization. One may ask how sensitive the proposed models are if these two parameters are varied; major change with the variation may hinder the generalization capability of the model.

We further run LBM and LBM-G (with the default parameter setting) and vary two parameters associated with citation trajectory classification -- activation period from 3-7 years, and peak threshold from $0.45-0.95$. For each value, we report the relative proportion as the ratio of the new proportion (say, $x$) with the new parameter value and the proportion (say, $y$) with the default parameter value, i.e., $\frac{x}{y}$. We measure this relative proportion for each category and for each model along with the real-world data. In Figure \ref{fig:mas_robust}, we observe that the change is insignificant ($p<0.005$) on the MAS dataset (same results are obtained for APS dataset shown in Figure \ref{fig:aps_robust}), hinting upon the fact that the conclusions drawn throughout the paper remain invariant with the minor change in the parameters related to trajectory categorization.

\section{Conclusion}
In this paper, we proposed two models to explain an important characteristic of a citation network -- the trajectory of citation growth. The models focus more on exploring the local neighborhood of node during edge formation, instead of looking at the network globally. This is important because edge formation in real networks is usually a local process. 
In typical network growth scenarios, nodes in the network either have limited information about the other nodes in the network or the system allows access to only restricted portion of the existing network. It therefore becomes imperative to model how the local processes of link formation gives rise to network characteristics. 
The proposed models do not consider any temporal factor (such as `aging') as a parameter, but still is able to realize various trajectories in citation growth.
Experimental results showed significant improvement  over the existing growth models on two real-world datasets in terms of realizing different citation trajectories of papers.

{\small
 \bibliographystyle{splncs04}
 \bibliography{references.bib}
}

\end{document}